\date{}
\documentstyle[preprint,eqsecnum,aps,epsf]{revtex}
\begin{document}
\title
{Characterization of Landau$-$Zener Transitions \\
in Systems with Complex Spectra}
\author{M. J. S\'anchez$^{a,b}$, E. Vergini$^{c}$ and D.A. Wisniacki$^{b}$.}
\maketitle

\noindent
\begin{center}
\center{$^{a}${\it Division de Physique Th\'eorique\footnote{Unit\'e
de recherche des Universit\'es de Paris XI et Paris VI associ\'ee au CNRS.

e-mail: majo@df.uba.ar},
Institut de Physique Nucl\'eaire. \\
91406, Orsay Cedex, France.}}\\
\center{$^{b}${\it Departamento de F\'{\i}sica, Facultad de Ciencias Exactas
 y Naturales, \\
Universidad de Buenos Aires.
Ciudad Universitaria, 1428 Buenos Aires, Argentina.}}\\ 
\center{$^{c}${\it Departamento de F\'{\i}sica, Comisi\'on Nacional de
Energ\'{\i}a At\'omica.\\ 
 Av. del Libertador 8250, 1429 Buenos Aires, Argentina.}}\\
\end{center}

\begin{abstract} 
This paper is concerned with  the study of one-body 
dissipation effects in idealized models resembling a nucleus. 
In particular, we study the quantum mechanics of a free particle that collides
elastically with the slowly moving walls of a Bunimovich stadium billiard.
Our results are twofold. First,
we develop a method to solve in a simple way  the quantum mechanical  evolution
of planar billiards with moving walls. The formalism is based
on the {\it scaling method} \cite{ver} which enables the resolution 
of the problem in terms of quantities defined over  the  boundary of 
the billiard.\\
The second result is related to the quantum aspects of dissipation in 
systems with complex spectra. We conclude 
that in a slowly varying evolution the e\-ner\-gy is transferred 
from the boundary to the particle through   Landau$-$Zener transitions.
\end{abstract}

\section*{Introduction}
\label{sec:int}
The way in which energy is transferred from the  time dependent 
mean field to the individual nucleons is an important ingredient, 
 for instance, in fission 
processes \cite{nix} and in large amplitude collective motion 
at low energies. Because of the Pauli principle it is expected 
that one body effects, {\it i.e.}, loss of energy due to collision of
independent individual nucleons with the mean field, should dominate the
dissipation mechanism.\\
Several descriptions of these processes  involving different approximations
are available in the literature \cite{swa1,hof,yan1,nor}.
These theories are perturbative in character and linear in the collective
motion. Therefore they are not suited to address the issues related to 
nonlinear dynamics and the onset of chaos.
On the other hand, the integrable or chaotic nature of the motion 
is of crucial importance to the dissipation mechanism \cite{swa2},{\it i.e.}, 
the transition from order to chaos 
provides the possibility for  a variety of nuclear responses (from elastic  to
elastoplastic to dissipative).
Therefore detailed studies of simplified models resembling a nucleus 
may be of interest.\\ 
Planar billiards are perhaps the best systems to model  the processes
described above in which the  nucleus can be imagined as a 
time-dependent container filled with a gas of non interacting point particles
 \cite{swa1,jarzi}.
Billiard systems have been thoroughly studied  in the context of classical 
and quantum chaos \cite{lh}.
In particular it has been shown that the quantum
spectra of generic planar billiards have GOE (Gaussian Orthogonal Ensemble)
characteristics that are  observed in the excited spectra of nuclei 
\cite{bo}.\\
In a seminal paper, Hill and Wheeler \cite{hill} suggested the 
Landau$-$Zener (L-Z) transitions as a mechanism for nuclear dissipation.
The mechanism is based on the  excitation of the 
individual nucleons via  transitions at  avoided 
level crossings near the Fermi surface. These excitations produce the damping 
of collective motion describing deformation of the nucleus.  
In an adiabatic evolution of the collective coordinates, the nucleus
changes its shape relatively slowly, while the nucleonic levels move 
up and down in energy. Small deformations in the nuclear shape  produce
occasionally that 
two nucleonic levels almost cross each other and experience an avoided
crossing. During the whole process many avoided crossings
occur with more or less random transitions between  nearest neigbours, in 
such a way that the system may end up in an arbitrary energy state. 
Within this picture one can imagine  a  stochastic dynamics in which 
by simply reversing the temporal evolution one does not recover the initial 
state. Therefore, the internal degrees of freedom are excited and the 
motion of the collective coordinates is thus damped.\\
More recently Wilkinson, making good use of the properties of complex
spectra, introduced a statistical treatment of dissipation in finite-sized quantum
systems in terms of L-Z transitions in the context of random 
matrix theory \cite{wil}. 
However, a L-Z mechanism as the generator of dissipation 
in systems with complex spectra, has recently been seriously 
questioned \cite{bulg}. In their works Bulgac and collaborators suggest 
that the diffusive process in energy
is dominated by memory effects and that the picture for dissipation 
through L-Z transitions is likely to be incorrect (see section \ref{3niv}
for a detailed analysis of these arguments).\\ 
Obviously the best way to elucidate this question is to solve the quantum 
mechanical evolution of a generic system, though this is difficult  
even for planar billiards with moving walls.\\

The goal of the present article is to  present a new
formulation to solve the quantum mechanical evolution of planar billiards 
with moving boundaries. Using this
formulation we  study the time evolution of a specific billiard, the
Bunimovich Stadium with  externally driven  walls.
We  restrict the analysis to slowly varying (adiabatic) evolutions 
\cite{landau}.
The notion of slow motion will be quantified in section \ref{fr}.
Our aim is to 
understand whether a L-Z mechanism generates the 
damping of the slow degree of freedom in a system with complex 
spectra.\\

The work  outline is as follows. In the next section we introduce a
one-dimensional new
formulation in order to solve the quantum 
evolution of planar billiards with moving walls .
Section \ref{numres} is devoted to the numerical results 
obtained for the Bunimovich Stadium
billiard with GOE spectrum characteristics. Using these results, we  
evaluate relevant properties for the dissipation process. 
In section \ref{3niv} we discuss in detail whether
 a L-Z transition mechanism describes the dissipation of 
the slow degree of freedom or equivalently the difussion of the fast ones.
The last section is devoted to final remarks and conclusions.\\
Before proceeding we want to stress that
planar billiards systems externally driven can also be used to model 
other problems often
encountered  for example in  mesoscopic systems, atomic
clusters and of course deformable cavities \cite{mesos}. Therefore, the 
 results presented in this work could  also contribute to domains other
than nuclear physics.

\section{The Method}
\label{sec:met}
\begin{figure}
\begin{center}
\leavevmode
\epsfysize=7in
\epsfbox{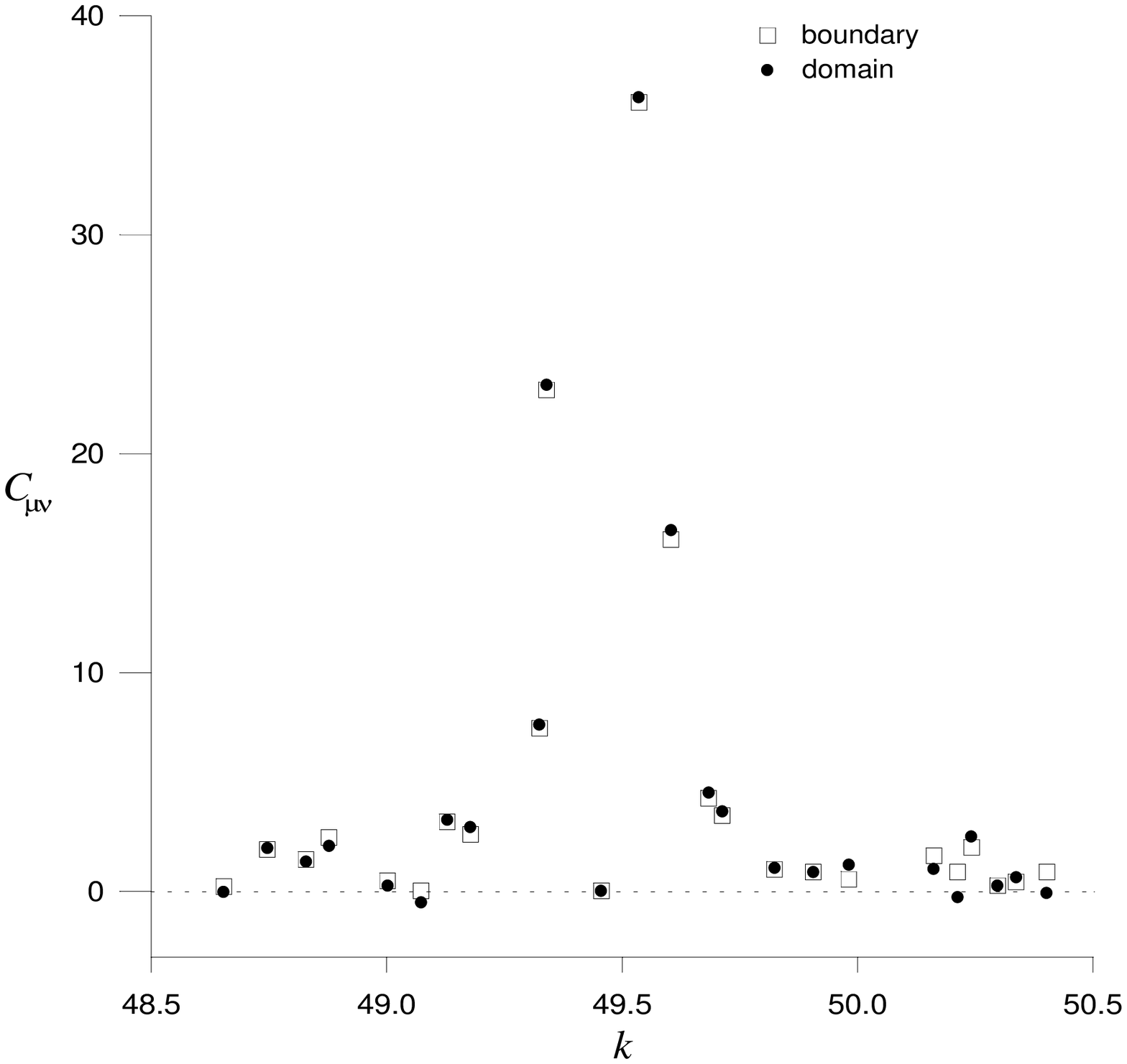}
\end{center}
\caption{Coefficients $C_{\mu \nu}$ (with $k_{\mu}=49.456279$)
as a function of $k_{\nu}$ computed 
exactly (equation (4)) and employing  the boundary definition 
(equation (8)). The system  used is
introduced in section \protect{\ref{numres}}
and the calculation corresponds to $t=0$.} 
\label{coe}
\end{figure}

In a recent work  Vergini and Saraceno  developed a method to
calculate directly {\it all} eigenvalues and eigenfunctions in a narrow energy 
range of quite general time independent $2-d$ billiards,  
by solving a generalized eigenvalue problem in terms of quantities defined
over the boundary.
The method is based on the use of scaling  
that enables to write the boundary norm explicitly as a function of the
energy  \cite{ver}.\\
The aim of this section is to extend the  
method of scaling    
to solve the Schr\"{o}dinger equation for $2\!-\!d$ billiards with time
dependent boundary conditions.\\
Let ${\cal C}(t)$ be a closed curve defining at time $t$ a two dimensional 
 domain ${\cal D}(t)$. We  restrict to star shaped domains, this means that 
$r_{n}\equiv {\bf r}.{\bf n} > 0 \; \forall {\bf r} \in {\cal C} (t); 
 \; {\bf n}$ is the outgoing normal to ${\cal C} (t)$.
Consider a particle of mass $m$ inside the billiard, then
the Schr\"{o}dinger equation reads, 

\begin{equation}
\label{scho}
{\partial{\Psi}\over\partial{t}} = i \frac{\hbar}{2m} \Delta \Psi \;.
\end{equation}
$\Psi$ satisfies the time dependent boundary condition 
$\Psi({\bf \zeta} ,t)= 0$  where ${\bf \zeta}$ is a point
on ${\cal C}(t)$, and we consider functions normalized to one
on the domain. A standard procedure is to expand the solution 
in terms of the adiabatic basis,

\begin{equation}
\label{base}
\Psi({\bf r},t)=\;\sum_{\mu} a_{\mu}(t) \; P_{\mu}(t) \; 
\phi_{\mu}({\bf r},t) \;.  
\end{equation}
$ \; P_{\mu}(t) \equiv \exp(-i\int_{0}^{t} 
\omega_{\mu} dt') \; $ is the contribution of the dynamical  phase 
with $\omega_{\mu}= \hbar k_{\mu}^{2}(t)/ \;2m$.
The adiabatic basis $\;\{\phi_{\mu}\}\;$ constitutes a complete set of real 
eigenfunctions of the  billiard at each time; that is, $\phi_{\mu}$
 satisfies the Helmholtz equation 
$\Delta\phi_{\mu}({\bf r},t)=- k_{\mu}^{2}(t) \phi_{\mu}({\bf r},t)$ 
 with  Dirichlet boundary condition $\phi_{\mu}({\bf \zeta},t)= 0$,
and it is a continuous function of time.\\
We  generate from the original domain defined by  ${\cal C} (t)$
a family of systems that depends on a parameter $\alpha$. These systems
evolve with the curves ${\cal C}_{\alpha} (t)$ that are obtained 
from ${\cal C} (t)$  through a scaling
transformation on the plane ${\bf r}\rightarrow \alpha {\bf r}$
(if ${\bf \zeta}$ is a point on ${\cal C} (t)$, then ${\bf \zeta} /\alpha$ 
is the corresponding point on ${\cal C}_{\alpha} (t)$).\\
To each function $\phi_{\mu}({\bf r}, t)$ we associate 
 the scaling function  $\phi_{\mu}(\alpha, {\bf r}, t)\equiv
 \phi_{\mu}(\alpha {\bf r}, t)$.
This family of functions depending on the scaling parameter $\alpha$ verifies
the Helmholtz equation with wave number $ {\alpha} k_{\mu}$ and satisfies the
Dirichlet condition over the scaled boundary.
Moreover, we require that the mass of the particle in the scaled systems
changes as ${\alpha}^{2} m$ in such a way that $\omega_{\mu}$ is 
independent of $\alpha$. The last statement implies that the time evolution
is the same for {\it all} the scaled systems.\\
Our approach to solve the Schr\"odinger equation is 
to expand  the wave function in terms of  the adiabatic basis
represented by the scaling functions. After replacing the expansion into
 the equation $(\ref{scho})$ we  obtain, 
\begin{equation}
\label{auto1}
\sum_{\nu} \dot{a}_{\nu}(t)  \;  P_{\nu}(t) \; \phi_{\nu}
( \alpha, {\bf r}, t) = \; -
 \sum_{\nu} a_{\nu}(t) \;  P_{\nu}(t) \; {\partial\phi_{\nu} 
\over\partial{t}}(\alpha, {\bf r}, t) \; .
\end{equation}
Differentiating this  equation with respect to $\alpha$ results in,

\begin{equation}
\label{der2}
\sum_{\nu} \dot{a}_{\nu}(t)  \;  P_{\nu}(t)\; {\partial\phi_{\nu}
\over\partial{\alpha}}
(\alpha, {\bf r}, t) = \; -
\sum_{\nu} a_{\nu}(t) \;  P_{\nu}(t) \; {\partial^{2}\phi_{\nu}\over
\partial{\alpha}\partial t } (\alpha, {\bf r}, t)\; .
\end{equation}
The remainder of the calculus consists on commuting the order of the 
partial derivation in the rhs of $(\ref{der2})$, multiply the equation by $
\; {\partial\phi_{\mu} / \partial{\alpha}} \; (\alpha, {\bf r}, t) \;$ and 
specialize 
the resulting equation in  $\alpha=1$. Finally we integrate over 
the boundary of the billiard ${\cal C} (t)$.
After this straightforward calculation, the final equation
reads, 

\begin{equation}
\label{final}
\sum_{\nu} \dot{a}_{\nu}(t)  \;  P_{\nu}(t) \;  \oint_{{\cal C}(t)} 
{\partial\phi_{\mu}
\over\partial{\alpha}} {\partial\phi_{\nu}
\over\partial{\alpha}} 
{dl\over r_{n}} 
= \; -
 \sum_{\nu} a_{\nu}(t) \;  P_{\nu}(t) \; \oint_{{\cal C}(t)} {\partial\phi_{\mu}
\over\partial{\alpha}} 
{\partial\over\partial{t}}({\partial\phi_{\nu}
\over\partial{\alpha}}) \; 
{dl\over r_{n}}  \;,
\end{equation}
where $dl$ is the length element on the boundary.
For the sake of simplicity we have omitted the argument  
$(\alpha = 1, {\bf r}, t)$ in the last equation.
In \cite{ver} it was proved  that the integral in the lhs of the 
last equation verifies a quasiorthogonality relation, this means

\begin{equation}
{1 \over 2  k_{\mu}^2} \oint_{{\cal C}(t)} {\partial\phi_{\mu}\over
\partial{\alpha}} 
{\partial\phi_{\nu}\over\partial{\alpha}} {dl\over r_{n}} =
 \delta_{\mu\nu} +{(k_{\mu}- k_{\nu}) \over (k_{\mu}+ k_{\nu})} 
{\cal O} (1) \;.
\end{equation} 
Employing this important relation  in (\ref{final}),
we derive the standard system of differential equations in the 
adiabatic basis
\begin{equation}
\label{auto2}
 \dot{a}_{\mu}(t) = \; -\sum_{\nu} a_{\nu}(t) ( P_{\nu}(t)/ P_{\mu}(t)) \; 
C_{\mu\nu}(t)
\end{equation}
with the coefficients $C_{\mu\nu}$ computed approximately in terms of
quantities defined over the boundary, 
\begin{equation}
\label{1cmunu}
 {\it C}_{\mu\nu}(t) \simeq {1\over 2 k_{\mu}^2}
\oint_{{\cal C}(t)} {\partial\phi_{\mu}
\over\partial{\alpha}} 
{\partial\over\partial{t}}({\partial\phi_{\nu}
\over\partial{\alpha}}) 
{dl\over r_{n}}  \;.
\end{equation}
The exact expression for the coefficients  follows from 
$(\ref{auto1})$,
\begin{equation}
\label{2cmunu}
C_{\mu\nu}(t)=
\int_{{\cal D}(t)} 
\phi_{\mu}({\bf r},t) \;{\partial\phi_{\nu}\over\partial{t}}({\bf r},t) \;   
d \sigma \;. 
\end{equation}
In the last equation each $C_{\mu\nu}(t)$ involves an
integration on the domain ${\cal D}(t)$. It is also very easy to prove
that they are antisymmetric.\\
To compute
each $C_{\mu \nu}$ via the equation $(\ref{2cmunu})$ the domain of integration
has to be partitioned at least
in $N \approx {k}^2$ cells, with $k$ equal to the maximum 
among the wave numbers  of the
functions $\phi$ in the region of integration, $k= max \{ k_{\mu},k_{\nu} \}$. 
If $n$ is the 
dimension of the adiabatic basis restricted to the energy region where
the evolution will take place, one needs  to know  $n^{2}$ coefficients.
As a consequence, the dimension of the problem 
of finding the coefficients $C_{\mu \nu}$ from equation $(\ref{2cmunu})$
turns to be of the order of $n^{2} \times N $ at each time.\\
As a way to check the goodness of the new formulation 
we have computed
for the specific billiard studied in section \ref{numres},  
the  coefficients ${\it C}_{\mu \nu}$$(t= 0)$ for a fixed $\mu$, with 
$ \nu=\mu+j$ $(j=0, \pm 1, \pm 2, ...)$  calculated exactly (equation 
(\ref{2cmunu})) and using the equation (\ref{1cmunu}) (see Fig.~\ref{coe}).
The correspondance is extremely good over a great number of levels.
The departure  between the two plots begins for
 $\mid j \mid \approx 10$, but in this region  the values of
the coefficients are very small.\\
With the present formulation the CPU   
 time necessary to compute the coefficients is considerably reduced in comparison
with the time needed in the standard approach (equation (\ref{2cmunu})).  
From the preceding remarks, and in order to 
study the interaction between neighbouring levels in the 
spectrum, we will calculate the coefficients $C_{\mu\nu}$ employing
the relation (\ref{1cmunu}).

\section{Numerical Results}
\label{numres}
\begin{figure}
\begin{center}
\leavevmode
\epsfysize=7in
\epsfbox{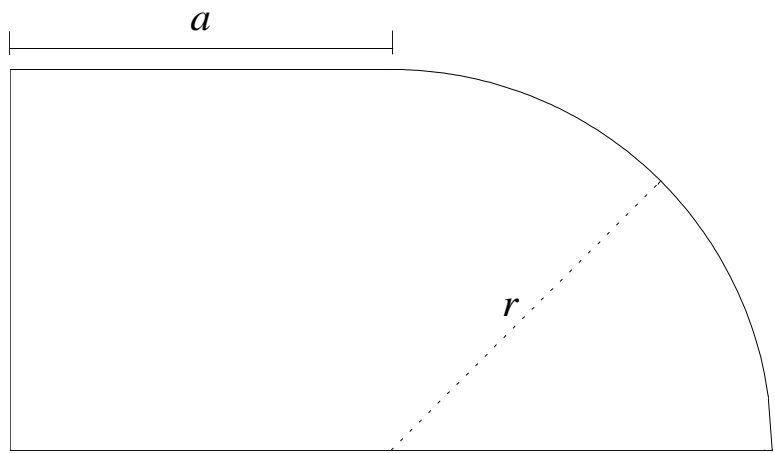}
\end{center}
\caption{Desymmetrized Bunimovich stadium billiard. The area of the billiard
is fixed to the value $1+\pi/4$. Then, the boundary only depends
on one parameter ($\ell \equiv a/r$).} 
\label{estadio}
\end{figure}

\begin{figure}
\begin{center}
\leavevmode
\epsfysize=7in
\epsfbox{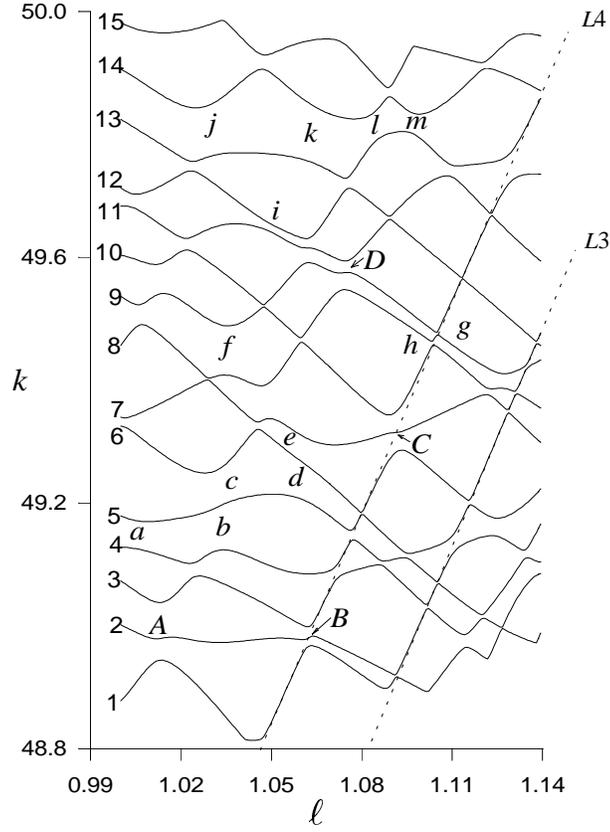}
\end{center}
\caption{Spectrum of the Bunimovich stadium billiard as a function 
of $\ell$, $1 \le \ell \le 1.14$. 
The wave numbers $k_{\mu} (\ell)$ run between $48.8$ and $50$. See 
text for more details.} 
\label{spectra}
\end{figure}

Using the method presented in section \ref{sec:met},
 we will analize the dynamics of a particle
of mass $m$ inside a Bunimovich stadium billiard with moving boundaries.\\ 
A point particle inside the static stadium billiard is a very well known
example of a fully classical chaotic system \cite{buni}. The particle 
moves  freely on the two dimensional domain and  is 
perfectly reflected  from its boundary.
The boundary is formed by 
two semi-circles of radius $r$ connected by two straight lines of 
length $2a$.
Fig.~\ref{estadio} shows a desymmetrized version of the system with
area $1+\pi/4$.\\
To study the dynamics, the parameter $\ell \equiv a/r$ is changed
with a finite velocity $\dot{\ell}$ in such a way that  the 
total area of the billiard remains unchanged. We have fixed the area
to avoid a drift  in the energy spectrum; this situation is
characteristic of nuclear processes where the nucleonic density is
approximately constant. The drift term represents a
reversible change in the energy of the system and  can be neglected 
in the analysis of an irreversible dissipation process \cite{wilki}.
Therefore, the dynamics 
of the boundary is introduced  through  the function $\ell (t)$.\\
Fig.~\ref{spectra} shows the spectrum of $k=\sqrt{2 m E}/\hbar$ as a function 
of $\ell$, $1 \le \ell \le 1.14$. We have selected  
the wave numbers $k_{\mu} (\ell)$  between $48.8$ and $50$ because 
in this region a large number of energy levels exists in a 
narrow portion of the spectrum. Although the properties that we 
are going to evaluate are characteristic of this region of the 
spectrum ($k \sim 50$), as we will show below, a proper 
scaling can be done in order to evaluate them in other energy regions.\\
The spectrum exhibits the typical behavior of avoided levels crossings that 
characterizes the energy levels as a function of a parameter for general 
systems without constants of motion \cite{berry}.
Also, we recognize that some avoided crossings are situated on two  parallel lines 
labeled  $L3$ and $L4$ (see Fig.~\ref{spectra}). 
These lines are associated to  bouncing ball states with three and
four low excitations respectively. 
These states are highly localized in the momenta space \cite{heller}, 
therefore  their interaction 
with  neighbouring  states  is smaller than  the
interaction  between delocalized generic states.\\

\begin{figure}
\begin{center}
\leavevmode
\epsfysize=7in
\epsfbox{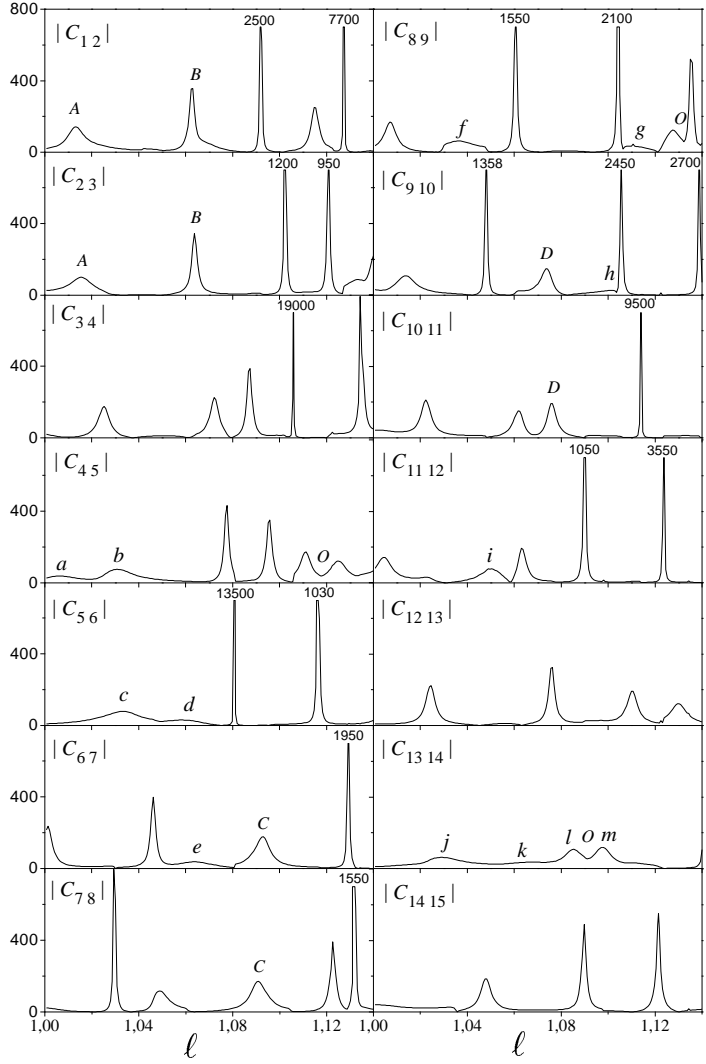}
\end{center}
\caption{$\mid \! C_{\mu \; \mu+1} \mid$ as a function of $\ell$ for 
several energy first neighbouring levels. The labels  {\it a, b, c..,} show 
small peaks that correspond to avoided crossings whose parameters can not 
be obtained directly from the spectrum. The peaks that 
correspond to interaction with bouncing ball states are out of scale and 
their maximum values are shown. The labels $A,\;B,\;C$ 
and $D$ show well defined
peaks that in the spectrum appears as  three levels avoided crossing.
The label $O$ shows  few cases where there is some overlap between
consecutive avoided crossings.} 
\label{cmunu}
\end{figure}

Let us analize the coefficients $C_{\mu \nu}$ which determine the quantum
mechanical evolution of the system (see section \ref{sec:met}).
They may be expressed
in terms of the deformation parameter $\ell$, as they satisfy 
$\; : C_{\mu \nu}(t) = \dot{\ell} C_{\mu \nu}(\ell)$.\\
Fig.~\ref{cmunu} shows 
the functions $\;\mid \!C_{\mu \mu+1} (\ell)\!\mid\;$ for several 
pairs of nearest neighbouring levels.  
A well defined structure of peaks is observed. 
The peaks appear  
each time  two neighbouring  energy levels experience an avoided crossing 
(it is very easy to follow in the Fig.~\ref{spectra} a pair of energy levels
as a function of the parameter $\ell$ in order to confirm this assertion).
The height of the peaks diminishes when  the  
energy gap between levels  at the avoided crossing increases. For
this reason the peaks corresponding to interaction with bouncing ball 
states are one order of magnitude greater than the generic ones.
We label with {\it a, b, c..,} small peaks that 
correspond to not well defined avoided crossings or to situations 
where  it is still difficult
to decide whether an avoided crossing exists by simple inspection of the 
spectrum (see also Fig.~\ref{spectra}).\\
For second neighbouring 
levels, we also find some well defined peaks, they appear essentially 
when three 
levels come close to each other (this situation is discussed in the next 
section). The heights of the  $C_{\mu \mu+2} (\ell)$ 
peaks are one order of magnitude smaller than those of the  $C_{\mu\mu+1}
(\ell)$.\\
For coefficients with $|\mu-\nu|>2$, we do not observe any simple structure,
however the amplitude of these coefficients is indeed  very small, lower 
than five in the scale of  Fig.~\ref{coe}, in comparison with the 
amplitude observed for nearest neighbouring coefficients.\\
From the present analysis it is clear that the information contained 
in the coefficients enables a complete definition of the avoided 
crossings and that this information is not always 
available in the spectrum.\\

\begin{figure}
\begin{center}
\leavevmode
\epsfysize=7in
\epsfbox{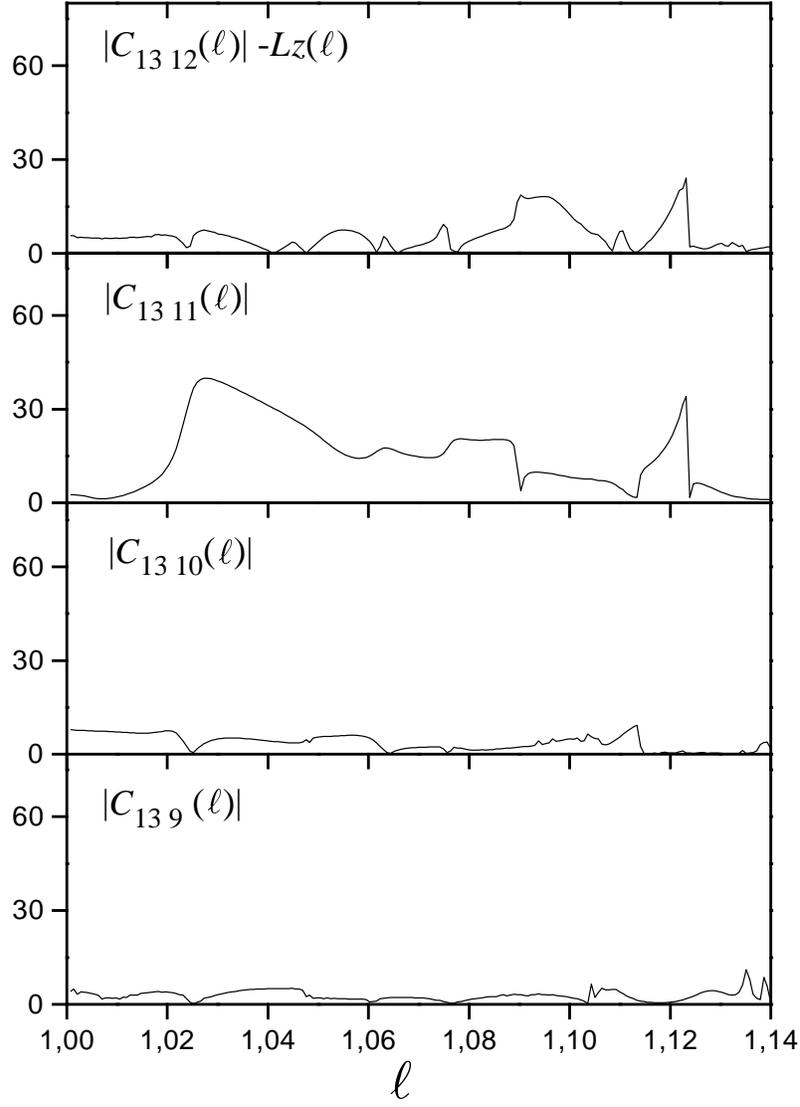}
\end{center}
\caption{The top shows $\mid \! C_{13 \; 12}(\ell) \mid -Lz(\ell)$ as a 
function $\ell$. $Lz(\ell)$ is a sum of lorenztian functions centered
on the peaks of the coefficient $\mid \! C_{13 \; 12}(\ell) \mid $. 
Their widths
and the position of the centers are $\ell_{int}$ and
$\ell_{0}$ respectively. The remainder part of the figure shows the
 coefficients $\mid \! C_{13 \; 13-j}(\ell) \mid $ for $j=2,3,4$ as a function
 of $\ell$.} 
\label{coef}
\end{figure}

The peaks between first neighbouring levels
are very well fitted by lorentzian functions, as it is expected for a
L-Z transition (the first part of the next section
is devoted to explain the expected lorentzian behavior of the coefficient), 
for almost all peaks with
the exception of some small ones.\\
Fig.~\ref{coef}  summarizes the preceding remarks.
It shows on the top the function $\mid C_{13 \; 12}(\ell)\mid  -Lz(\ell)$
where $Lz(\ell)$ is a sum of lorenztian functions centered on the well
defined peaks of the coefficient $C_{13 \; 12}(\ell)$. Each lorentzian function
is defined  in the next section by the equation $(\ref{c21})$. The widths
 and the position of the centers are $\ell_{int}$ and
 $\ell_{0}$ respectively.\\
The remainder of the Fig.~\ref{coef} shows the functions 
$\mid C_{13 \; 13-j}(\ell) \mid $
$(j=2,3,4)$ as a function of $\ell$. The figure reveals the lorentzian
behavior of the first neighbouring levels coefficients, and the lack
of a defined structure in the coefficients $C_{\mu \nu} (\ell)$ 
for $|\mu-\nu|>2$.\\

\begin{figure}
\begin{center}
\leavevmode
\epsfysize=7in
\epsfbox{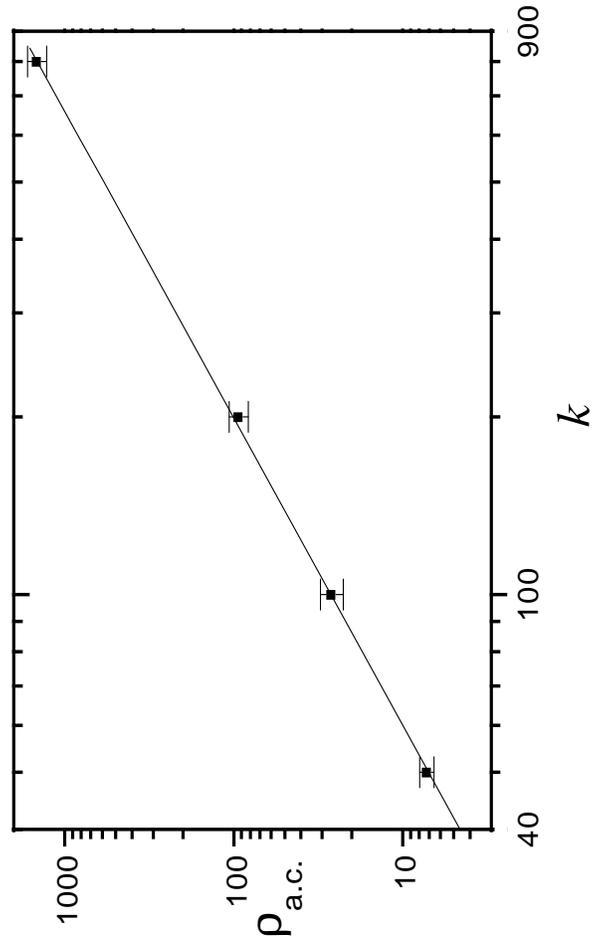}
\end{center}
\caption{Log-Log plot of density of avoided crossings $\rho_{a.c}$ as a 
function of $k$ for gap sizes less than one quarter of the mean level 
spacing.   $\rho_{a.c}$   scales as $k^{d}$ with $d=1.92\pm0.1$.}
\label{plot}
\end{figure}

The previous numerical study would be still more appealing if we knew
how the spectrum scales to other energy regions. The Weyl's law
\cite{baltes}
tells us  that the density of states associated to the  vertical axis in
Fig.~\ref{spectra}, scales as $k$. The problem appears with the horizontal
axis because the scaling of the density of consecutive avoided crossings
 $\rho_{a.c}$ is unknown.\\
Working in different energy regions and after an exhaustive numerical 
analysis, we have obtained that  $\rho_{a.c}$ scales as 
$k^{d}$ with $d=1.92\pm0.1$. Fig.~\ref{plot} shows  $\rho_{a.c}$ as a 
function of $k$ for gap sizes less than one quarter of the mean level 
spacing. In this calculation we have not
considered avoided crossings with bouncing ball states because their
relative contribution to the density of states decreases 
as $k^{-1/2}$ \cite{berry2}.\\
Another important fact to stress is that each peak is very well defined;
its width (given by $\ell_{int}$) is much smaller than  the mean distance
between consecutive peaks ${\rho_{a.c}}^{-1}$.
Only in few cases a small overlap between consecutive avoided 
crossings is observed (see for example the peaks labeled by {\it O} 
in Fig.~\ref{cmunu}).
For generic avoided crossings we have obtained that 
$\ell_{int} \; {\rho_{a.c}} \approx 0.2$; for avoided crossings with
bouncing ball states this product is even smaller as expected.\\

\section{ Landau$-$Zener Behavior}
\label{3niv}
As we have mentioned in the introduction, our aim
is to understand whether L-Z transitions govern the mechanism of one
body dissipation in systems with complex spectra.\\
In the previous section we have analyzed  the coefficients $C_{\mu\nu}$ 
that describe the quantum evolution of a particular system with complex 
spectrum.
The analysis revealed that the coefficients have a simple structure
of well defined lorentzian peaks as the dominant contribution, plus
a very small component without any defined structure (see Fig.~\ref{coef}).
These peaks are concentrated in the first neighbouring
levels  coefficients, which, as  we will show below, is a characteristic of L-Z
transitions.
In few cases well defined peaks appear in coefficients
between second neighboring levels, but these peaks are one order
of magnitude smaller than the previous ones. We will discuss this
situation below using an idealized three level system.\\

We begin this section with a brief review on the theory of L-Z transitions.
Consider a two level system that depends  on 
a parameter $\ell$, in such a way that for ${\ell} = {\ell}_{0}$ 
the energy
levels experience an avoided crossing. Let  $\phi_{\pm}(\ell)$   
be the adiabatic eigenstates and 
$\;E_{+}-E_{-}= \sqrt{\gamma^{2}(\ell-\ell_{0})^2 + 4 \epsilon^{2} }\;$
the energy gap between the associated eigenvalues, with $\gamma$ 
and $\epsilon$ constants. 
The adiabatic theorem \cite{boh} tells us that if the system is initially 
in the state
 $\phi_{-}$ and $\ell$ changes infinitely slowly from  ${\ell} \le 
 {\ell}_{0} $ to  $ {\ell} \ge {\ell}_{0}$ 
the system will remain in the state $\phi_{-}$. However, if $\ell$ changes
with a finite velocity  the final state will be 
a linear combination of the basis states. 
Zener derived the  probability of an adiabatic transition 
employing the diabatic basis \cite{diab} for a constant velocity of 
the parameter  ${\dot\ell}$.
If at time
$t= -\infty$ the system were in the state $\phi_{-}$ the transition probability 
at time $t=\infty$ is 
$P_{z}= exp( -2\pi {\epsilon}^2 / \gamma \dot\ell \hbar )$ 
 \cite{zener}.\\
Using the adiabatic basis and (\ref{2cmunu}) is straightforward to derive
\begin{equation}
\label{c21}
C_{+-}(\ell)= { \ell_{int} \over 2 (\ell_{int} ^2 + (\ell-\ell_{o})
^2)} \; \;,
\end{equation}
where  $\ell_{int} \equiv 2  \epsilon / \gamma $
is the width of the
lorentzian function; that is, the characteristic time for a L-Z 
process.\\

\begin{figure}
\begin{center}
\leavevmode
\epsfysize=7in
\epsfbox{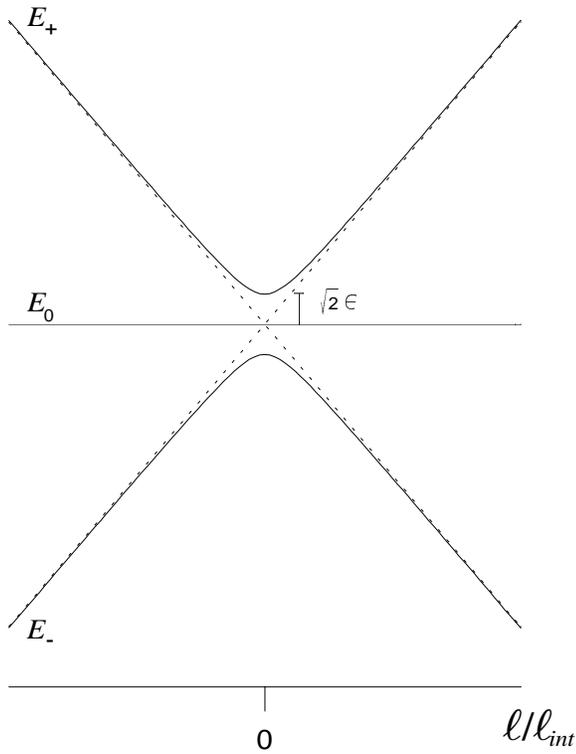}
\end{center}
\caption{Energy levels $E_{+}$,
 $E_{-}$ and $E_{0}$ as a function of 
$\ell/ \ell_{int}$
for the $3-$levels hamiltonian analized in section
\protect{\ref{3niv}}. 
The dotted lines indicates 
the asymptotes to $E_{-}$ and $E_{+}$. The distance
 $(E_{+}-E_{0})(0)=
\protect{\sqrt{2}} 
\epsilon$ 
is also drawn.} 
\label{3nive}
\end{figure}

It has been suggested  in  recent literature that it is very difficult
to characterize the interaction between neighbouring
levels in spectra like the present one in terms of L-Z transitions
\cite{bulg}.
The arguments could be summarized as follows: i) It is not always possible
from the spectrum to localize the position of the avoided crossings and to
determine the parameters that define the L-Z transitions.
This assertion is partially true; it is not in the spectrum where all
the information is contained. 
We have solved this problem employing the adiabatic basis, in which the 
position of the avoided crossing and the interaction length $\ell_{int}$ 
are well defined in terms of the coefficients $C_{\mu\nu}$.\\
ii) The L-Z transition probability  is 
exponentially small when the length of the transition process goes to 
infinity, but it could be strongly affected for lengths of the
 order of $\ell_{int}$ \cite{balian}. This problem could emerge if the mean 
distance between avoided crossings $\rho_{a.c}$ is of the order or less 
than $\ell_{int}$.
In the preceding section
we have obtained $\ell_{int} \; {\rho_{a.c}} \approx  0.2$ for generic avoided
crossings (between delocalized eigenfunctions) and
this value is highly reduced for localized eigenfunctions. In a physical 
system the eigenfunctions present some degree of localization because the
associated classical phase space is not fully chaotic. Therefore, we do not
expect correlations between consecutive avoided crossings.
In terms of the coefficients
$C_{\mu \mu+1}$, the last assertion means that each individual peak is very 
well defined.\\
iii) One often encounters a situation where three levels come close to each 
other and by simple inspection one can think in a three level crossing 
(see for example points $ A,\; B,\; C$ and $D$ in the Fig.~\ref{spectra}).
In order to understand this process we will analyze a three level system 
which mimics such a circumstance.\\
Consider a one parameter dependent Hamiltonian defined in the diabatic basis
by the following matrix, 
\[
\epsilon \; \left( \begin{array}{ccc}
            -\ell / \ell_{int} &  1 & 0 \\
      1 & 0         & 1 \\
       0       & 1  & \ell / \ell_{int} 
\end{array}
\right)
\]
with $\epsilon$ the perturbation and $\ell_{int}$ the characteristic transition
length. This Hamiltonian can be diagonalized analytically for each time.
The upper and lower eigenenergies are represented by hyperbolas
$E_{\pm}=  \pm \epsilon \sqrt{(\ell / \ell_{int})^2 +2 } \;$, as in the
L-Z
process, and the middle energy is $E_{0}=0$ for all times (see 
Fig.~\ref{3nive}).
Obviously, for a diabatic evolution, if the system were in the upper state at 
$-\ell / \ell_{int}  >> 1$, there is a high probability $\sim 1$ that 
the system decays to the
state $E_{-}$ at $\ell / \ell_{int}  >> 1$. 
In other words, the presence of $E_{-}$ affects
enormously the transition probability between $E_{+}$ and $E_{0}$. However, we
are interested in an adiabatic evolution where the transition 
probability to $E_{-}$
turns to be small. In such a situation we want to determine whether 
the L-Z parameters
for $E_{+}$ and $E_{0}$, that is $(E_{+}-E_{0})/2$ at  $\ell=0$ 
(we denote it $\Delta$) and $\ell_{int}$, adequately
describe the transition probability between these two states. This point 
is not obvious at all. For 
example, the  distance $(E_{+}-E_{0})$ is largely affected by the presence of
$E_{-}$ and  $\Delta \ne \epsilon$, contrary to the case of a two level
system.\\
We have computed numerically the dynamical evolution of this model  
obtaining the 
following result for the transition probability between $E_{+}$ and $E_{0}$:
\begin{equation}
P_{E_{+}\rightarrow E_{0}} \sim {1\over 2} \exp[- \pi \; \Delta
\; (\ell_{int}/\dot\ell \hbar) \; 0.96]
\end{equation}
for $P_{E_{+}\rightarrow E_{0}} \lesssim 0.2$. That is, although the parameters 
need to be renormalized, the factor is very close to $1$. As a
conclusion, this  three levels system
may be thought as two independent  avoided crossings
with L-Z interactions.

\section{Final Remarks}
\label{fr}
The results of this paper attempt to extend the present understanding of
one body dissipation processes.
To analyze the way in which energy is transferred form the time dependent mean
field to the individual nucleons  we have modeled
the mean field by a slowly time dependent container. The same approach has been
employed by many other authors in related nuclear models \cite{swa2,jarzi}.
The container is represented by a planar billiard with externally driven
moving walls.
To solve the quantum mechanical evolution of these simplified
systems, we have derived a one-dimensional formulation. This approach gives the
possibility to study the evolution of highly excited states and
reduces the CPU time involved in the calculations.\\
We have devoted part of the work to answer a fundamental question;
whether a Landau$-$Zener excitation mechanism governs the irreversible
transport of energy from the driven wall to the particles in an adiabatic
evolution.
We have analyzed a parameter dependent billiard system with GOE character
spectrum, concluding that in an adiabatic evolution  of the external parameter,
the dissipation is dominated by  L-Z transitions at the
avoided crossings.
The {\it adiabatic limit} is attained in the limit of an infinitely slow
evolution. On the other hand, adiabatic evolution refers to 
slowly varying evolutions \cite{landau}. Of course, the notion of slow motion
needs to be clarified. For example, we have excluded in our analysis
the structure showed by 
the function $C_{13 \; 11}(\ell)$ in Fig.~\ref{coef} because its height
is very small; although the area
under it is comparable to the area under any peak 
observed in Fig.~\ref{cmunu}. However, because $C_{13 \; 11}(\ell)$ is 
multiplied  in the differential equation (\ref{auto2})
 by an
oscillatory function with period $T \approx \hbar \rho_{E}$ ($\rho_{E}$ is
the density of energy levels), its effective contribution is canceled 
if the time required by
the collective motion to sweep the structure 
$\;t_{coll} \approx \rho_{a.c}^{-1}/ \dot{\ell}\;$ is larger than $T$.\\
The above adiabaticity condition is satisfied by systems where
quantum effects are very important, such as nuclei. However, as the 
wave number increases, the collective velocity needs to be 
reduced drastically.
Taking the semiclassical limit $\hbar \rightarrow 0$,
$k \rightarrow \infty $, with $\;(k \hbar)^{2}/2m =E=const\;$,
it results that
$\;T={\cal O}(k^{0})\;$ and $\;t_{coll}={\cal O}(k^{-d}/\dot{\ell})\;$.
Therefore, for any finite  
value of $\dot{\ell}$, the evolution is always {\bf diabatic} in the 
semiclassical limit. In another terms, a semiclassical theory
of dissipation requieres  a  scaling of $\dot{\ell}$. 
To our knowledge, this important point has not been taken 
into account in  previous works \cite{bulg}.\\

The description of the damping process in terms of L-Z
transitions has been already done by Wilkinson in the context of pure
random matrix theory \cite{wil,wilki} but to our knowledge the present work is
the first study carried out for a  more realistic system. If
the L-Z behavior holds it is
more or less straigthforward to write the corresponding diffusion equation to
quantify the dissipation mechanism (for further details  see \cite{wilki}).

\section*{Acknowledgments}
We specially thank the hospitality of the Center Emile Borel
where part of this work has been done. E.V acknowledges support from
the Service de Physique Th\'eorique CEA-Saclay.\\
We would like to thank O.Bohigas and P. Leboeuf for useful suggestions.\\
M.J.S is supported by a Thalmann fellowship of the Universidad de Buenos Aires,
Argentina, and is on leave from CONICET, Argentina.

\newpage

\newpage

\newpage

\newpage

\newpage

\newpage

\newpage

\newpage
 
\end{document}